\documentclass[12pt]{article}
\usepackage{amsmath}
\usepackage{graphicx}
\usepackage{amsfonts}
\usepackage{amssymb}
\newtheorem{theorem}{Theorem}

\newtheorem{proposition}[theorem]{Proposition}

\begin{document}

\author{John Miritzis\\Department of Marine Sciences, University of the Aegean\\Electronic address: john@env.aegean.gr}
\title{Dynamical system approach to FRW models in higher-order gravity theories}
\date{\today}
\maketitle
\begin{abstract}
We study the late time evolution of positively curved FRW models with a scalar
field which arises in the conformal frame of the $R+\alpha R^{2}$ theory. The
resulted three-dimensional dynamical system has two equilibrium solutions
corresponding to a de Sitter space and an ever expanding closed universe. We
analyze the structure of the first equilibrium with the methods of the center
manifold theory and, for the second equilibrium we apply the normal form
theory to obtain a simplified system, which we analyze with special phase
plane methods. It is shown that an initially expanding closed FRW spacetime
avoids recollapse.
\end{abstract}

\section{Introduction}

A central question in relativistic cosmology is that of deciding about the
past and future asymptotic states of cosmological models \cite{wael}. However,
general relativity leads to singularities in the spacetimes of all known
cosmological models with ordinary matter. Higher order curvature corrections
in the gravitational action may rectify the problem and lead to cosmological
models free from such pathologies, at the cost of diverging from a FRW
behavior at late times \cite{ruzm}. There is a resurgence of interest in such
theories which naturally arise in string-theoretic considerations (cf. brane
models with Gauss-Bonett terms \cite{lno,chdu,muko,abmo}). An interesting
feature of higher order theories is that inflation emerges in these theories
in a most direct way. In one of the first inflationary models, proposed in
1980 by Starobinsky \cite{star}, inflation is due to the $R^{2}$ correction
term in a gravitational Lagrangian $L=R+\alpha R^{2}$ where $\alpha$ is a
constant. The dynamics of higher order cosmologies is closely related to
scalar-field cosmologies in general relativity because of their conformal
equivalence \cite{ba-co88,gmss}. There are certain limitations to this
procedure related to the issue of physical reality of the two metrics involved
\cite{cots} and to the fact that the conformal transformation may fail to be
regular at all points of the spacetime. Nevertheless, it is practically useful
and investigations in the conformal frame have given some interesting results,
e.g. the cosmic no-hair theorem in quadratic cosmologies \cite{maed,comi}.

Most of the studies of scalar-field cosmologies with the dynamical systems
methods are restricted to FRW models (see for example \cite{guetal,cole} and
references therein), although there are important investigations in spatially
homogeneous Bianchi cosmologies with an exponential potential \cite{coibho}.
In particular, for flat FRW models with a scalar field there are some general
results which do not rely on the particular properties of the potential
\cite{fost,miri}. However, the situation is more delicate for positively
curved FRW models with a scalar field having a potential with a zero local
minimum. The main problem which confronts us is the following: Can a closed
universe filled with ordinary matter and a scalar field avoid recollapse?

In this paper we investigate the evolution of positively curved FRW models
with a scalar field having the potential which arises in the conformal frame
of the $R+\alpha R^{2}$ theory in vacuum \cite{ba-co88,maed}. The motivation
for this choice is presented in Section III. The dimension of the dynamical
systems involved in such models is greater than two and the usual methods of
phase plane analysis cannot be applied. In particular, for nonhyperbolic
equilibrium points the linearization theorem does not yield any information
about the stability of the equilibria and therefore, more powerful methods are
needed. The center manifold theorem shows that the qualitative behavior in a
neighborhood of a nonhyperbolic equilibrium point $\mathbf{q}$ is determined
by its behavior on the center manifold near $\mathbf{q\,.}$ Since the
dimension of the center manifold is generally smaller than the dimension of
the dynamical system, this greatly simplifies the problem. The other general
method for simplifying the dynamical system is the normal form theory, which
consists in a nonlinear coordinate transformation that allows to simplify the
nonlinear part of the system. Both methods are used in Sections III and IV respectively.

The plan of the paper is as follows. In the next Section we write down the
field equations assuming an arbitrary potential, as a constrained
four-dimensional dynamical system. In Section III, we use the constraint
equation to reduce the dimension of the system to three and after a suitable
change of variables we find the equilibrium points of the system and discuss
the physical meaning of these particular solutions. In particular we show
using the methods of the center manifold theory that, the equilibrium
corresponding to the de Sitter solution is asymptotically unstable. In Section
IV we find the so-called normal form of the dynamical system, which greatly
simplifies the problem, since two of the equations decouple. We study the
qualitative behavior of the resulted two-dimensional system and analyze the
late time evolution of the model. We show that an initially expanding universe
avoids recollapse. In Section V we apply the same techniques to flat FRW
spaces filled with a barotropic fluid in the conformal frame of the $R+\alpha
R^{2}$ theory and study the detailed evolution of the models.

\section{Scalar-field cosmologies}

In general relativity the evolution of FRW models with a scalar field
(ordinary matter is described by a perfect fluid with energy density $\rho$
and pressure $p$) are governed by the Friedmann equation,
\begin{equation}
\left(  \frac{\dot{a}}{a}\right)  ^{2}+\frac{k}{a^{2}}=\frac{1}{3}\left(
\rho+\frac{1}{2}\dot{\phi}^{2}+V\left(  \phi\right)  \right)  , \label{fri1jm}%
\end{equation}
the Raychaudhuri equation,
\begin{equation}
\frac{\ddot{a}}{a}=-\frac{1}{6}\left(  \rho+3p+2\dot{\phi}^{2}-2V\right)  ,
\label{fri2jm}%
\end{equation}
the equation of motion of the scalar field,
\begin{equation}
\ddot{\phi}+3\frac{\dot{a}}{a}\dot{\phi}+V^{\prime}\left(  \phi\right)  =0
\label{emsjm}%
\end{equation}
and the conservation equation,
\begin{equation}
\dot{\rho}+3\left(  \rho+p\right)  \frac{\dot{a}}{a}=0. \label{conssfjm}%
\end{equation}
We adopt the metric and curvature conventions of Ref. \cite{wael}. Here,
$a\left(  t\right)  $ is the scale factor, an overdot denotes differentiation
with respect to time $t$ and units have been chosen so that $c=1=8\pi G.$

From Eqs. (\ref{fri1jm})-(\ref{conssfjm}) we see that the state $\left(
a,\dot{a},\rho,\phi,\dot{\phi}\right)  \in\mathbb{R}^{5}$ of the system lies
on the hypersurface defined by the constraint, (\ref{fri1jm}), and the
remaining field equations can be written as a five-dimensional dynamical
system. In vacuum, $\rho=0,$ the dimension of the dynamical system reduces to four.

In the literature of scalar-field cosmologies the exponential potential
function \textit{viz}. $V\left(  \phi\right)  =V_{0}e^{-\lambda\phi}$, is the
most popular not only because of the variety of alternative theories of
gravity which predict exponential potentials, but also due to the fact that
this potential has the nice property that $V^{\prime}\propto V$ which allows
the introduction of normalized variables according to the formalism of
Wainwright \textit{et} \textit{al} \cite{wael}. In flat, $k=0$ FRW models for
example with a scalar field having an exponential potential, introducing the
variables $x\sim\dot{\phi}/H,\;y\sim\sqrt{V}/H$ and the time coordinate
$\tau=\ln\left(  a/a_{0}\right)  $, enables the evolution equations to be
written as a two-dimensional dynamical system (cf. \cite{coliwa}) and in more
general homogeneous cosmologies associated with a scalar field, the dimension
of the dynamical system reduces by one if the potential function is exponential.

If we set $\dot{\phi}=y,\,\;\dot{a}/a=H$ the evolution equations
(\ref{fri2jm})-(\ref{conssfjm}) in vacuum become%
\begin{align}
\dot{a}  &  =Ha\nonumber\\
\dot{\phi}  &  =y,\nonumber\\
\dot{y}  &  =-3Hy-V^{\prime}\left(  \phi\right)  ,\nonumber\\
\dot{H}  &  =-\frac{1}{2}y^{2}+k/a^{2}, \label{k1}%
\end{align}
subject to the constraint
\begin{equation}
3H^{2}+3k/a^{2}=\frac{1}{2}y^{2}+V\left(  \phi\right)  . \label{constk}%
\end{equation}
Therefore, the phase space of the dynamical system (\ref{k1}) is the set
\[
\left\{  \left(  a,\phi,y,H\right)  \in\mathbb{R}^{4}:3H^{2}+3k/a^{2}=\frac
{1}{2}y^{2}+V\left(  \phi\right)  \right\}  .
\]

\section{Curved FRW in $R+\alpha R^{2}$ theory: Equilibria}

In the remainder of the paper we assume that the potential function of the
scalar field is
\begin{equation}
V\left(  \phi\right)  =V_{\infty}\left(  1-e^{-\sqrt{2/3}\phi}\right)  ^{2}
\label{pote}%
\end{equation}
which arises in the conformal frame of the $R+\alpha R^{2}$ theory
\cite{ba-co88,maed}. This potential has a long and flat plateau. For large
values of $\phi,$ the potential, $V,$ is almost constant, $V_{\infty}%
=\lim_{\phi\rightarrow+\infty}V\left(  \phi\right)  ,$ thus $V$ has the
general properties for inflation to commence. In \cite{comi} it was proved a
cosmic no-hair theorem, i.e. Bianchi models with ordinary matter satisfying
the strong energy condition and a scalar field with potential (\ref{pote}),
asymptotically isotopize. According to this picture, the universe started in a
homogeneous state and during inflation it had enough time to isotropize.

In order to reduce the dimension of the dynamical system (\ref{k1}) we use the
constraint (\ref{constk}) to eliminate $a.$ The evolution equations become%
\begin{align}
\dot{\phi}  &  =y,\nonumber\\
\dot{y}  &  =-3Hy-V^{\prime}\left(  \phi\right)  ,\nonumber\\
\dot{H}  &  =-H^{2}-\frac{1}{3}y^{2}+\frac{1}{3}V\left(  \phi\right)  .
\label{sys1}%
\end{align}
Linearization of (\ref{sys1}) near the equilibrium point $\left(
0,0,0\right)  $ shows that the Jacobian matrix at that point has one zero and
two purely imaginary eigenvalues. Consequently the Hartman-Grobman theorem
does not apply. Therefore, we cannot draw any conclusions about the stability
of the equilibrium from an examination of the Jacobian.

We simplify the system by rescaling the variables by the equations%
\begin{align*}
\phi &  \rightarrow\sqrt{2/3}\phi,\\
y &  \rightarrow\sqrt{4V_{\infty}/3}\,\,y,\\
H &  \rightarrow\frac{\sqrt{2V_{\infty}}}{3}H,\\
t &  \rightarrow\frac{1}{\sqrt{2V_{\infty}}}t.
\end{align*}
In order to take account of the equilibrium point corresponding to the point
at ``infinity'' and to remove the transcendental functions, it is convenient
to introduce the variable $u$ defined by
\begin{equation}
u:=e^{-\phi},\label{u}%
\end{equation}
to obtain finally
\begin{align}
\dot{u} &  =-uy,\nonumber\\
\dot{y} &  =-Hy-u\left(  1-u\right)  ,\nonumber\\
\dot{H} &  =-\frac{1}{3}H^{2}-\frac{2}{3}y^{2}+\frac{1}{2}\left(  1-u\right)
^{2}.\label{sys3}%
\end{align}
Note that under the transformation (\ref{u}), the resulted three-dimensional
dynamical system (\ref{sys3}) is quadratic. In view of (\ref{constk}) we have
$3H^{2}-\frac{1}{2}y^{2}-V\left(  \phi\right)  >0,$ hence, the phase space of
the system (\ref{sys3}) is the set
\begin{equation}
\Sigma:=\left\{  \left(  u,y,H\right)  \in\mathbb{R}^{3}:H^{2}-y^{2}-\frac
{3}{2}\left(  1-u\right)  ^{2}>0\right\}  .\label{ineq1}%
\end{equation}

The equilibrium points of (\ref{sys3}) are:

A: $\left(  u=1,y=0,H=0\right)  .$ This corresponds to the limiting state of
an ever-expanding universe with $H\rightarrow0$ while the scalar field
approaches the minimum of the potential and the scale factor goes to infinity.
Equality in (\ref{ineq1}) which arises from the flat, $k=0,$ case defines a
set on the boundary of $\Sigma.$ We conclude that the point A which
corresponds to the Minkowski solution, is located on this boundary. The
detailed structure of this equilibrium will be analyzed in the next Section.

B: $\left(  u=0,y=\pm\sqrt{3}/2,H=0\right)  .$ These lie outside of the phase
space and, therefore, are unphysical.

C: $\left(  u=0,y=0,H=\pm\sqrt{3/2}\right)  .$ In the next Section we show
that only the point with the $+$ sign can be approached by a trajectory
starting with a $H>0.$ It corresponds to the de Sitter universe with a
cosmological constant equal to $\sqrt{V_{\infty}}$. Regarding the stability of
this equilibrium, it is easy to see that the Jacobian matrix of (\ref{sys3})
at $\mathbf{q}=\left(  0,0,\sqrt{3/2}\right)  $ has one zero and two negative
eigenvalues. The center manifold theorem implies that there exists a local
2-dimensional stable manifold through $\mathbf{q}$ (see for example
\cite{perko}). That means that all trajectories asymptotically approaching
$\mathbf{q}$ as $t\rightarrow\infty,$ lie on a 2-dimensional invariant
manifold. Since $\mathbf{q}$ is a nonhyperbolic fixed point, the topology of
the flow near $\mathbf{q}$ is nontrivial and is characterized by a
one-dimensional local center manifold intersecting $\mathbf{q}.$ In the
Appendix we prove the following result.

\begin{proposition}
The equilibrium point $\mathbf{q}=\left(  0,0,\sqrt{3/2}\right)  $ of
(\ref{sys3}) is locally asymptotically unstable.
\end{proposition}

\section{Late time evolution}

It is easy to see that at the equilibrium point, $\left(  u=1,y=0,H=0\right)
,$ the eigenvalues of the Jacobian of (\ref{sys3}) are $\pm i,0$ and
therefore, we cannot infer about the stability of the equilibrium.
Nevertheless, it is the most interesting case, because in all other equilibria
the scalar field reaches the flat plateau, which is impossible if we restrict
ourselves to initial values of $H$ smaller than $\sqrt{V_{\infty}}$. The study
of the qualitative behavior of a dynamical system near a nonhyperbolic
equilibrium point is difficult even in two dimensions. We find the so-called
normal form (cf. \cite{perko} for a brief introduction) of the system
(\ref{sys3}) near the equilibrium point $\left(  u=1,y=0,H=0\right)  .$ The
idea of the normal form theory is the following: Given a dynamical system with
equilibrium point at the origin, $\mathbf{\dot{x}}=A\mathbf{x}+\mathbf{f}%
\left(  \mathbf{x}\right)  ,$ where $A$ is the Jordan form of the linear part
and $\mathbf{f}\left(  \mathbf{0}\right)  =\mathbf{0}$, perform a non-linear
transformation $\mathbf{x}\rightarrow\mathbf{x}+\mathbf{h}\left(
\mathbf{x}\right)  ,$ where $\mathbf{h}\left(  \mathbf{x}\right)  =O\left(
\left|  \mathbf{x}\right|  ^{2}\right)  $ as $\left|  \mathbf{x}\right|
\rightarrow0,$ such that the system becomes ``as simple as possible''.

To write the system in a form suitable for the application of the normal form
theory, we shift the fixed point to $\left(  0,0,0\right)  $ by setting
$x=u-1$ and the system becomes%
\begin{align}
\dot{x}  &  =-y-xy,\nonumber\\
\dot{y}  &  =x+x^{2}-Hy,\nonumber\\
\dot{H}  &  =\frac{1}{2}x^{2}-\frac{2}{3}y^{2}-\frac{1}{3}H^{2}. \label{sys4}%
\end{align}
We now perform the non-linear transformation
\begin{align*}
x  &  \rightarrow x-y^{2}+\frac{1}{4}Hy,\\
y  &  \rightarrow y+xy+\frac{1}{4}Hx,\\
H  &  \rightarrow H+\frac{7}{12}xy,
\end{align*}
and keeping only terms up to second order we obtain the system
\begin{align}
\dot{x}  &  =-y-\frac{1}{2}Hx,\nonumber\\
\dot{y}  &  =x-\frac{1}{2}Hy,\nonumber\\
\dot{H}  &  =-\frac{1}{12}\left(  x^{2}+y^{2}\right)  -\frac{1}{3}H^{2}.
\label{cartesian}%
\end{align}
Note that the results are valid only near the origin.

Passing to cylindrical coordinates $\left(  x=r\cos\theta,y=r\sin
\theta,H=H\right)  ,$ we have
\begin{align}
\dot{r}  &  =-\frac{1}{2}rH,\;\nonumber\\
\;\dot{\theta}  &  =1,\nonumber\\
\dot{H}  &  =-\frac{1}{12}r^{2}-\frac{1}{3}H^{2}\;. \label{normal3d}%
\end{align}
We note that the $\theta$ dependence of the vector field has been eliminated,
so that we can study the system on the $r-H$ plane. The equation $\dot{\theta
}=1$ means that the trajectory in the $x-y$ plane spirals with angular
velocity $1.$ It is convenient to rescale the variables by%
\begin{equation}
r\rightarrow6r,\;\;H\rightarrow3H \label{resc}%
\end{equation}
so that the projection of (\ref{normal3d}) on the $r-H$ plane is
\begin{align}
\dot{r}  &  =-\frac{3}{2}rH,\;\nonumber\\
\dot{H}  &  =-r^{2}-H^{2}\;. \label{2dim}%
\end{align}
This system belongs to a family of systems studied in 1974 by Takens
\cite{takens}.

It is easy to obtain the phase portrait of (\ref{2dim}) via numerical
integration. However, we can analyze the qualitative behavior of the
trajectories using theoretical arguments. Firstly, (\ref{2dim}) is invariant
under the transformation $t\rightarrow-t,$ $H\rightarrow-H$ (which implies
that all trajectories are symmetric with respect to the $r$ axis) and the line
$r=0$ is invariant. Secondly, the system (\ref{2dim}) has invariant lines
$H=cr$. To see this, write%
\begin{equation}
\frac{dH}{dr}=c=\frac{-r^{2}-c^{2}r^{2}\;}{-\frac{3}{2}cr^{2}}=\frac
{-1-c^{2}\;}{-\frac{3}{2}c}\Rightarrow c=\pm\sqrt{2}. \label{inva}%
\end{equation}

Taking the dot product of the vector field $\left(  -\frac{3}{2}%
rH,-r^{2}-H^{2}\right)  ^{\perp}$ with the radial vector $\left(  r,H\right)
^{\perp}$ along the line $H=cr$ we find that it is negative for $H>0$ and
positive for $H<0.$ Therefore, the direction of the flow along $H=cr$ in the
first quadrant is towards the origin and goes away from the origin in the
second quadrant. Note that $H$ is always decreasing along the orbits while $r$
is decreasing in the first quadrant. Since no trajectory can cross the line
$H=cr,$ all trajectories starting above this line, approach the origin
asymptotically. On any orbit starting in the first quadrant below the line
$H=cr,$ $H$ becomes zero at some time and the trajectory crosses vertically
the $r-$axis. Once the trajectory enters the second quadrant, $r$ increases
and $H$ decreases. The phase portrait is shown in Figure 1.
\begin{figure}[h]
\begin{center}
\includegraphics[
trim=0.000000in 5.121122in 0.000000in 1.675067in, height=2.7726in,
width=4.6596in ]{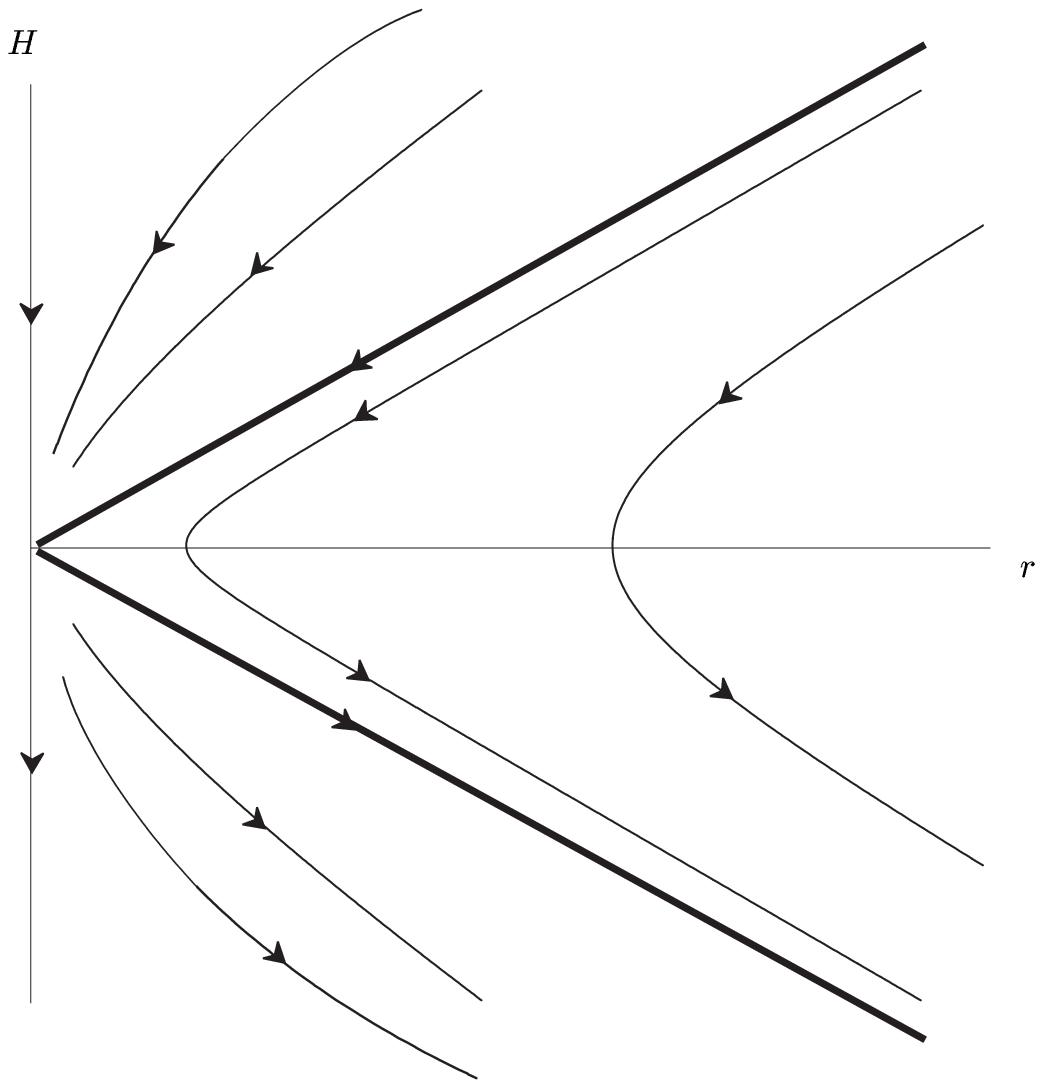}\newline
\end{center}
\caption{The phase portrait of (\ref{2dim}). The invariant
trajectories $H=\pm\sqrt{2}r$ are shown with thicker lines. For
expanding models only trajectories above the line $H=\sqrt{6}r$
belong to the phase space of the
system.}%
\label{fig1}%
\end{figure}
At first sight, it seems probable that an initially expanding universe may
recollapse. However, the phase space of the dynamical system (\ref{2dim}) is
not the whole $r-H$ plane, because of the constraint (\ref{ineq1}), which in
terms of the variables (\ref{resc}) becomes%
\begin{equation}
H^{2}>6r^{2}. \label{ineq2}%
\end{equation}
Therefore, for an expanding universe we should consider only trajectories
starting above the line $H=\sqrt{6}r$ and according to the previous discussion
all these trajectories asymptotically approach the origin.

We now turn to the relation of the dynamics in the $r-H$ plane to the full
three-dimensional system (\ref{normal3d}), or the equivalent (\ref{cartesian})
in Cartesian coordinates. Any trajectory spirals clockwise in the $x-y$ plane
while both $H$ and $x^{2}+y^{2}$ are decreasing. In physical terms this means
that the scalar field oscillates around the minimum of the potential with a
decreasing amplitude, $H$ is always decreasing and, in view of (\ref{constk}),
the curvature decreases. A typical trajectory of (\ref{cartesian}) is shown in
Figure 2.
\begin{figure}[h]
\begin{center}
\includegraphics[
trim=0.000000in 2.020141in 0.000000in 5.824136in, height=2.5702in,
width=5.4872in ]{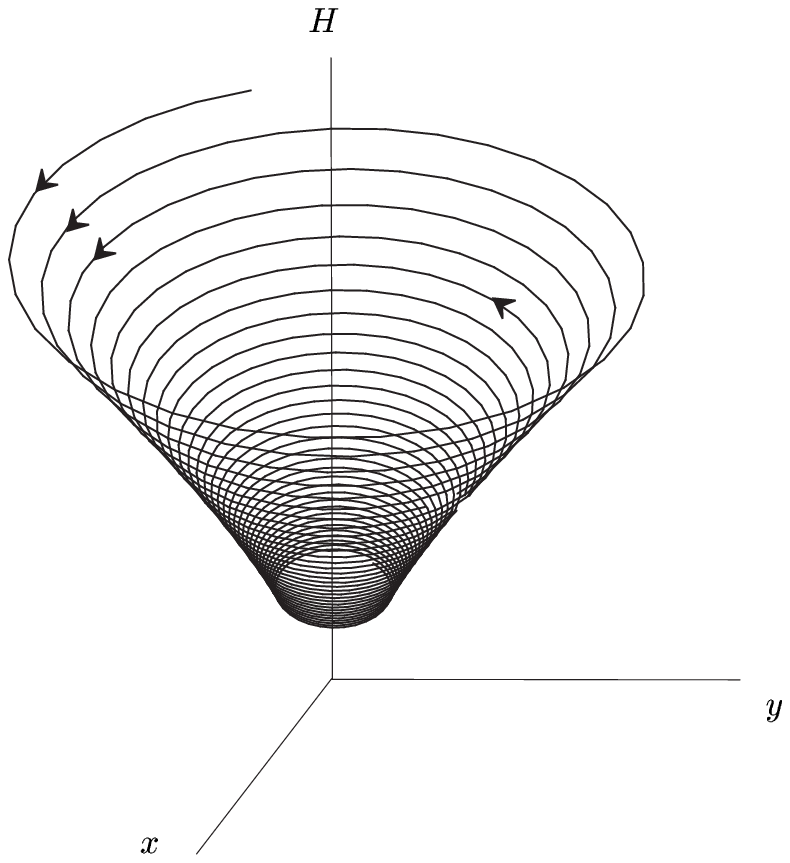}\newline
\end{center}
\caption{Trajectories of the full three-dimensional system spiral
approaching
the origin.}%
\label{fig2}%
\end{figure}
\section{Comment on Flat FRW models}

Consider a flat FRW model containing a barotropic fluid with an equation of
state $p=(\gamma-1)\rho,$ where $0\leq\gamma\leq2,$ and a scalar field having
the potential (\ref{pote}). Then the system (\ref{fri2jm})-(\ref{conssfjm})
reduces again to a four-dimensional dynamical system, namely%
\begin{align}
\dot{\phi} &  =y,\nonumber\\
\dot{y} &  =-3Hy-V^{\prime}\left(  \phi\right)  ,\nonumber\\
\dot{\rho} &  =-3\gamma\rho H,\nonumber\\
\dot{H} &  =-\frac{1}{2}y^{2}-\frac{\gamma}{2}\rho,\label{sfjm}%
\end{align}
subject to the constraint
\begin{equation}
3H^{2}=\rho+\frac{1}{2}y^{2}+V\left(  \phi\right)  .\label{costrsfjm}%
\end{equation}
In contrast to (\ref{k1}d), the forth of (\ref{sfjm}) implies that $H$ is
always decreasing. If we use the constraint (\ref{costrsfjm}) to eliminate
$\rho,$ the evolution equations become%
\begin{align}
\dot{\phi} &  =y,\nonumber\\
\dot{y} &  =-3Hy-V^{\prime}\left(  \phi\right)  ,\nonumber\\
\dot{H} &  =-\frac{3\gamma}{2}H^{2}-\frac{2-\gamma}{4}y^{2}+\frac{\gamma}%
{2}V\left(  \phi\right)  .\label{sy1}%
\end{align}
We see that the structure of (\ref{sy1}) describing a flat FRW model with a
perfect fluid and a scalar field, apart from being parameter dependent, has a
striking similarity to the dynamical system (\ref{sys1}) for the vacuum
positively curved FRW with a scalar field. Proceeding as in Section IV, we end
up with the following system in cylindrical coordinates
\begin{align}
\dot{r} &  =-\frac{1}{2}rH,\;\nonumber\\
\;\dot{\theta} &  =1,\nonumber\\
\dot{H} &  =\frac{5\gamma-4}{8}r^{2}-\frac{\gamma}{2}H^{2}\;.\label{nor3d}%
\end{align}
Although the system (\ref{nor3d}) depends only on one parameter, it is
convenient to rescale the variables by%
\begin{equation}
r\rightarrow\lambda r,\;\;H\rightarrow\mu H\label{resc1}%
\end{equation}
with%
\[
\mu=\frac{2}{\gamma},\;\;\lambda=\sqrt{\left|  \frac{16}{\gamma\left(
5\gamma-4\right)  }\right|  },
\]
so that the projection of (\ref{nor3d}) on the $r-H$ plane is
\begin{align}
\dot{r} &  =-\frac{1}{\gamma}rH,\;\nonumber\\
\dot{H} &  =br^{2}-H^{2}\;,\label{2di}%
\end{align}
where
\[
b=+1\;\mathrm{for\;}\gamma>4/5,\;\;b=-1\;\mathrm{for\;}\gamma<4/5.
\]

Note that (\ref{2di}) has a first integral, \textit{viz}.%
\begin{equation}
I\left(  r,H\right)  =-\frac{1}{2\gamma}r^{-2\gamma}\left(  \frac{br^{2}%
}{1-1/\gamma}-H^{2}\right)  .
\end{equation}
In fact, it is straightforward to verify that $\left(  \partial
I/\partial r\right)  \dot{r}+\left(  \partial I/\partial H\right)
\dot{H}=0$ along the solution curves of (\ref{2di}). The level
curves of $I$ are the trajectories of the system  (cf. Figure 3).

Invariant lines $H=cr$ exist for certain values of the parameter $\gamma.$ We
find (cf. (\ref{inva}))%
\[
c=\pm\sqrt{\frac{b}{1-1/\gamma}}.
\]

Case I. For $b=+1,$ invariant lines exists if $\gamma>1.$ We find that the
direction of the flow along $H=cr$ in the first quadrant is towards the origin
and goes away from the origin in the second quadrant. Note that in the first
quadrant $r$ is decreasing along the orbits and that $\dot{H}$ vanishes along
the line $H=r,$ which lies below the invariant line $H=cr.$ It can be shown
that in the first quadrant a level curve of $I\left(  r,H\right)  $ may
intersect the line $H=r$ only once (it is sufficient to consider the level
curve passing through an arbitrary point $\left(  r_{1},0\right)  $ and
compute the $r$ coordinate at the intersection with the line $H=r$). We
conclude that once a trajectory crosses the line $H=r,$ it is trapped between
the lines $H=r$ and $H=cr$ and, since $\dot{r}<0,$ it approaches the origin asymptotically.

Case II. $b=+1$ and $4/5<\gamma<1.$ There are no invariant lines. Similar
arguments as in case I, yield the phase portrait shown in Fig.3.

Case III. $\gamma<4/5$ $\left(  \Rightarrow b=-1\right)  .$ The analysis is
exactly the same as in Section IV.
\begin{figure}[h]
\begin{center}
\includegraphics[
trim=0.000000in 2.353517in 0.000000in 6.387950in, height=1.6527in,
width=4.5766in ]{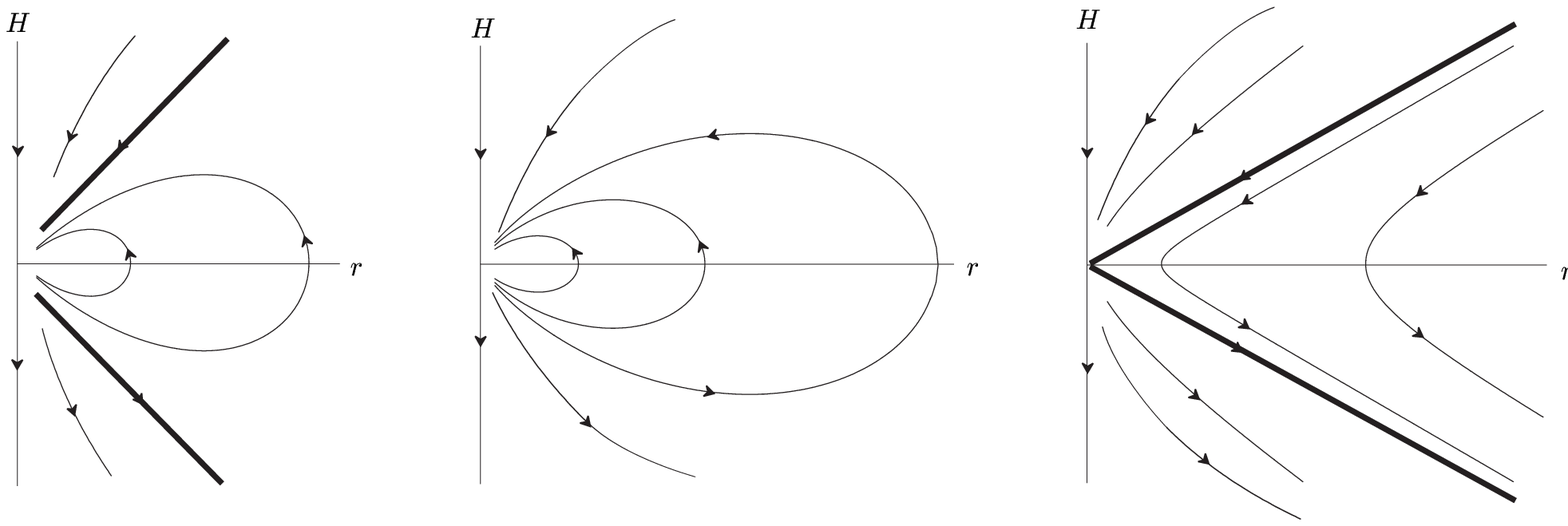}\newline
\end{center}
\caption{The phase portrait of (\ref{2di}) for $1<\gamma<2,$
$4/5<\gamma<1$
and $0<\gamma<4/5$ respectivelly.}%
\label{fig3}%
\end{figure}
In all cases we must remember that the phase space of the dynamical system
(\ref{2di}) is not the whole $r-H$ plane, because of the constraint
(\ref{ineq1}), which in terms of the variables (\ref{resc1}) reads%
\begin{equation}
H^{2}>\frac{6\gamma}{\left|  5\gamma-4\right|  }r^{2}.
\end{equation}
Therefore, for an expanding universe, we should consider only trajectories
starting above the line $H=\sqrt{6\gamma/\left|  5\gamma-4\right|  }r,$ which
in case III lies always above the line $H=cr$ and according to the previous
discussion all these trajectories asymptotically approach the origin.

We conclude that, in the conformal frame of the $R+aR^{2}$ theory, an
initially expanding flat universe with a barotropic fluid as matter source
remains ever-expanding and eventually the quadratic curvature corrections
become negligible. This result, established by stability analysis, is in
accordance with the general properties of all flat FRW models with a scalar
field having a potential with a unique zero minimum \cite{miri}.

\section{Discussion}

We have analyzed the qualitative behavior of a positively curved FRW model
containing a scalar field with the potential (\ref{pote}). This model is
conformally equivalent to the positively curved FRW spacetime in the simplest
higher order gravity theory, namely the $R+\alpha R^{2}$ theory. We have shown
that an initially expanding closed universe avoids recollapse provided that
the initial value of $H$ is less than $\sqrt{V_{\infty}}.$ This result should
be compared to a counterexample of the closed universe recollapse conjecture
(cf. \cite{cflq} where it is shown that initially expanding vacuum diagonal
Bianchi IX models in purely quadratic gravity are ever-expanding). It should
be of interest to investigate if a closed FRW universe filled with ordinary
matter satisfying the usual energy conditions and a scalar field with the
potential (\ref{pote}) can avoid recollapse. This is equivalent to the
analysis of the qualitative behavior of the full five-dimensional system
(\ref{fri2jm})-(\ref{conssfjm}). A partial answer to this question for an
arbitrary non-negative potential having a unique minimum $V\left(  0\right)
=0$ is given in \cite{miri}.

\bigskip

\textbf{Acknowledgments} I thank P.G.L. Leach and S. Cotsakis for fruitful
discussions during the preparation of this work. I wish to thank an anonymous
referee for useful suggestions, especially for pointing me out the remarks
regarding the equilibrium A after eq. (\ref{ineq1}).%

%TCIMACRO{\TeXButton{TeX field}{\setcounter{equation}{0}}}%
%BeginExpansion
\setcounter{equation}{0}%
%EndExpansion%
%TCIMACRO{\TeXButton{TeX field}{\renewcommand{\theequation}{A.\arabic
%{equation}}}}%
%BeginExpansion
\renewcommand{\theequation}{A.\arabic
{equation}}%
%EndExpansion
\appendix

\section{Proof of Proposition 1}

In order to determine the local center manifold of (\ref{sys3}) at
$\mathbf{q},$ we have to transform the system into a form suitable for the
application of the center manifold theorem. The procedure is fairly systematic
and will be accomplished in the following steps.

1. The Jacobian of (\ref{sys3}) at $\mathbf{q}=\left(  0,0,\sqrt{3/2}\right)
$ has eigenvalues $0,-\sqrt{2/3}$ and $-\sqrt{3/2}$ with corresponding
eigenvectors $\left(  -\sqrt{2/3},2/3,1\right)  ^{^{\top}},$ $\left(
0,0,1\right)  ^{^{\top}}$ and $\left(  0,1,0\right)  ^{^{\top}}.$ Let $T$ be
the matrix having as columns these eigenvectors. We shift the fixed point to
$\left(  0,0,0\right)  $ by setting $\widetilde{H}=H-\sqrt{3/2}$ and write
(\ref{sys3}) in vector notation as
\begin{equation}
\mathbf{\dot{z}}=A\mathbf{z}+\mathbf{F}\left(  \mathbf{z}\right)  ,
\label{sys3a}%
\end{equation}
where $A$ is the linear part of the vector field and $\mathbf{F}\left(
\mathbf{0}\right)  =\mathbf{0}$.

2. Using the matrix $T$ which transforms the linear part of the vector field
into Jordan canonical form, we define new variables, $\left(  x,y_{1}%
,y_{2}\right)  \equiv\mathbf{x}$, by the equations%
\begin{align*}
u  &  =-\sqrt{\frac{2}{3}}x,\\
y  &  =\frac{2}{3}x+y_{2},\\
\widetilde{H}  &  =x+y_{1},
\end{align*}
or in vector notation $\mathbf{z}=T\mathbf{x},$ so that (\ref{sys3a}) becomes
\[
\mathbf{\dot{x}}=T^{-1}AT\mathbf{x}+T^{-1}\mathbf{F}\left(  T\mathbf{x}%
\right)  .
\]
Denoting the canonical form of $A$ by $B$ we finally obtain the system
\begin{equation}
\mathbf{\dot{x}}=B\mathbf{x}+\mathbf{f}\left(  \mathbf{x}\right)  ,
\label{sys3b}%
\end{equation}
where $\mathbf{f}\left(  \mathbf{x}\right)  :=T^{-1}\mathbf{F}\left(
T\mathbf{x}\right)  .$ In components system (\ref{sys3b}) is%
\begin{align}
\left(
\begin{array}
[c]{c}%
\dot{x}\\
\dot{y}_{1}\\
\dot{y}_{2}%
\end{array}
\right)   &  =\left(
\begin{array}
[c]{ccc}%
0 & 0 & 0\\
0 & -\sqrt{\frac{2}{3}} & 0\\
0 & 0 & -\sqrt{\frac{3}{2}}%
\end{array}
\right)  \left(
\begin{array}
[c]{c}%
x\\
y_{1}\\
y_{2}%
\end{array}
\right) \nonumber\\
&  +\left(
\begin{array}
[c]{l}%
-\frac{2}{3}x^{2}-xy_{2}\\
\frac{10}{27}x^{2}-\frac{1}{3}y_{1}^{2}-\frac{2}{3}y_{2}^{2}-\frac{2}{3}%
xy_{1}+\frac{1}{9}xy_{2}\\
\frac{4}{9}x^{2}-\frac{2}{3}xy_{1}-\frac{1}{3}xy_{2}-y_{1}y_{2}%
\end{array}
\right)  . \label{sys3c}%
\end{align}

3. The system (\ref{sys3c}) is written in diagonal form%
\begin{align}
\dot{x}  &  =Cx+f\left(  x,\mathbf{y}\right) \nonumber\\
\mathbf{\dot{y}}  &  =P\mathbf{y}+\mathbf{g}\left(  x,\mathbf{y}\right)  ,
\label{sys3d}%
\end{align}
where $\left(  x,\mathbf{y}\right)  \in\mathbb{R}\times\mathbb{R}^{2},$ $C$ is
the zero $1\times1$ matrix, $P$ is a square matrix with negative eigenvalues
and $f,\mathbf{g}$ vanish at $\mathbf{0}$ and have vanishing derivatives at
$\mathbf{0.}$ The center manifold theorem asserts that there exists a
1-dimensional invariant local center manifold $W^{c}\left(  \mathbf{0}\right)
$ of (\ref{sys3d}) tangent to the center subspace (the $\mathbf{y}=\mathbf{0}$
space) at $\mathbf{0}.$ Moreover, $W^{c}\left(  \mathbf{0}\right)  $ can be
represented as
\[
W^{c}\left(  \mathbf{0}\right)  =\left\{  \left(  x,\mathbf{y}\right)
\in\mathbb{R}\times\mathbb{R}^{2}:\mathbf{y}=\mathbf{h}\left(  x\right)
,\;\left|  x\right|  <\delta\right\}  ;\;\;\;\mathbf{h}\left(  0\right)
=0,\;D\mathbf{h}\left(  0\right)  =\mathbf{0},
\]
for $\delta$ sufficiently small (cf. \cite{perko}, p. 155). The restriction of
(\ref{sys3d}) to the center manifold is
\begin{equation}
\dot{x}=Cx+f\left(  x,\mathbf{h}\left(  x\right)  \right)  . \label{rest}%
\end{equation}
According to Theorem 3.2.2 in \cite{guho}, if the origin $x=0$ of (\ref{rest})
is stable (resp. unstable) then the origin of (\ref{sys3d}) is also stable
(resp. unstable). Therefore, we have to find the local center manifold, i.e.,
the problem reduces to the computation of $\mathbf{h}\left(  x\right)  .$

4. Substituting $\mathbf{y}=\mathbf{h}\left(  x\right)  $ in the second
component of (\ref{sys3d}) and using the chain rule, $\mathbf{\dot{y}%
}=D\mathbf{h}\left(  x\right)  \dot{x}$, one can show that the function
$\mathbf{h}\left(  x\right)  $ that defines the local center manifold
satisfies%
\begin{equation}
D\mathbf{h}\left(  x\right)  \left[  Cx+f\left(  x,\mathbf{h}\left(  x\right)
\right)  \right]  -P\mathbf{h}\left(  x\right)  -\mathbf{g}\left(
x,\mathbf{h}\left(  x\right)  \right)  =0. \label{h}%
\end{equation}
This condition allows for an approximation of $\mathbf{h}\left(  x\right)  $
by a Taylor series at $x=0.$ Since $\mathbf{h}\left(  0\right)  =0,D\mathbf{h}%
\left(  0\right)  =\mathbf{0},$ it is obvious that $\mathbf{h}\left(
x\right)  $ commences with quadratic terms. We substitute%
\[
\mathbf{h}\left(  x\right)  =:\left(
\begin{array}
[c]{c}%
h_{1}\left(  x\right) \\
h_{2}\left(  x\right)
\end{array}
\right)  =\left(
\begin{array}
[c]{c}%
a_{1}x^{2}+a_{2}x^{3}+O\left(  x^{4}\right) \\
b_{1}x^{2}+b_{2}x^{3}+O\left(  x^{4}\right)
\end{array}
\right)
\]
into (\ref{h}) and set the coefficients of like powers of $x$ equal to zero to
find the unknowns $a_{1},b_{1},...$.

5. Since $y_{1}$ is absent from the first of (\ref{sys3c}), we give only the
result for $h_{2}\left(  x\right)  .$ We find $b_{1}=\frac{4}{9}\sqrt{\frac
{2}{3}},$ $b_{2}=\frac{4}{81}.$ Therefore, (\ref{rest}) yields%
\begin{equation}
\dot{x}=-\frac{2}{3}x^{2}-\frac{4}{9}\sqrt{\frac{2}{3}}x^{3}-\frac{4}{81}%
x^{4}+O\left(  x^{5}\right)  . \label{rest1}%
\end{equation}
It is obvious that the origin $x=0$ of (\ref{rest1}) is asymptotically
unstable (saddle point). The theorem mentioned after (\ref{rest}) implies that
the origin $\mathbf{x}=\mathbf{0}$ of the full three-dimensional system is
unstable. This completes the proof.

\end{document}